\documentclass[12pt,]{article}
\usepackage{amsfonts}
\usepackage{amsmath,amssymb,bm}

\input{tcilatex}

\begin{document}

\title{Some exact solutions with torsion in 5-D Einstein-Gauss-Bonnet gravity%
}
\author{F. Canfora$^{1,2}$, A. Giacomini$^1$, S. Willison$^1$ \\
$^1$ Centro de Estudios Cientificos (CECS), Casilla 1469 Valdivia, Chile.\\
$^2$ Istituto Nazionale di Fisica Nucleare, Sezione di Napoli, GC Salerno.\\
e-mail: canfora@cecs.cl, giacomini@cecs.cl, steve@cecs.cl}
\maketitle

\begin{abstract}
Exact solutions with torsion in Einstein-Gauss-Bonnet gravity are derived.
These solutions have a cross product structure of two constant curvature
manifolds. The equations of motion give a relation for the coupling
constants of the theory in order to have solutions with nontrivial torsion.
This relation is not the Chern-Simons combination. One of the solutions has
a $AdS_2\times S^3$ structure and is so the purely gravitational analogue of
the Bertotti-Robinson space-time where the torsion can be seen as the dual
of the covariantly constant electromagnetic field.
\end{abstract}

Keywords: Einstein-Gauss-Bonnet gravity, torsion. \newline
PACS: 04.50.+h, 04.20.Jb, 04.40.Nr \newline
\newline
Preprint: CECS-PHY-07/11

\section{Introduction}

It is a well known fact that in four dimensions the Einstein-Hilbert action
(plus a cosmological term) is the only functional that can be built out of
the curvature invariants leading to second order field equations. In fact
higher order terms in the curvature invariants lead generically to higher
order field equations which at quantum level would lead to ghosts spoiling
the unitarity of the theory\cite{Zwiebach:1985uq}. The Einstein Hilbert
action can however be generalized in a straightforward way to higher
dimensions. Indeed there exist large class of theories containing higher
powers in the curvature which lead to second order equations for the metric,
known as Lovelock theories \cite{Lovelock}.

Now in the Einstein-Hilbert action, the vielbein $e^{a}$ and spin connection 
$\omega ^{ab}$ can be treated as independent fields. This is known as the
first order formalism since the field equations involve only first
derivatives: such a formalism is mandatory when dealing with Fermionic
fields. One of the characteristic features of the Einstein-Hilbert action
(in $n$ dimensions) in the first order formalism is that its variation with
respect to the spin connection gives equations of motion of the form 
\begin{equation*}
\epsilon _{a_{1}\ldots a_{n}}T^{a_{1}}\wedge e^{a_{2}}\wedge \cdots \wedge
e^{a_{n-2}}=0
\end{equation*}%
which simply imply the vanishing of the torsion two-form $T^{a}=0$. However
in dimension higher than four the Einstein-Hilbert action is no longer the
unique possible first order action. In fact, the Lovelock theories also
admit a first order formulation \cite{Zumino:1985dp} \footnote{%
As well as the Lovelock terms, one can also add to the action terms
explicitly involving the torsion and Lorentz Chern-Simons terms related to
the Pontryagin form \cite{Mardones:1990qc}. However, in this paper we focus
on five dimensions, where no such terms exist.}. In five dimensions for
example one can add to the standard Einstein-Hilbert action with a
cosmological constant an extra term to the usual action known as the
Gauss-Bonnet term which in familiar notation reads 
\begin{equation*}
\int d^{5}x\sqrt{g}\left( R^{2}-4R_{\mu \nu }R^{\mu \nu }+R_{\mu \nu \rho
\sigma }R^{\mu \nu \rho \sigma }\right) .
\end{equation*}%
Above $R_{\mu \nu \rho \sigma }$ is the Riemann tensor, $R_{\mu \nu }$ the
Ricci tensor and $R$ the Ricci scalar. We shall employ the differential form
notation. Introducing the curvature two-form, $R^{ab}$, the above term is
equal to 
\begin{equation*}
\int R^{ab}\wedge R^{cd}\wedge e^{e}\,\epsilon _{abcde}.
\end{equation*}%
Besides being a natural generalization of General Relativity to the five
dimensional case, Einstein-Gauss-Bonnet theories appear to be quite
compatible with the available astrophysical and cosmological experimental
data (see, for an incomplete list of references, \cite%
{CEN06,CN06,GOT06,GTS05,KM06,LN07, NO07,NOS05,NOS06a,STT05,San06,SB06}). The
addition of Lovelock terms to the action affects the equations of motion in
such a way that they no longer imply that the torsion vanishes. Instead, it
becomes a new propagating degree of freedom. The presence of torsion could
have interesting phenomenological consequence (see, for instance, Ref. \cite%
{AD01}). Torsion has a deep geometrical meaning which could shed some light
on non-perturbative features of gravitational theories which cannot be taken
into account in the standard formalism.

Generally the Lovelock equations of motion, combined with the Bianchi
identities, give very strong constraints on the torsion \cite%
{Troncoso:1999pk}. In most cases one obtains an over-determined system of
equations, making it extremely difficult to find exact solutions with
non-vanishing torsion. There is a special case where solutions with torsion
are known: in odd dimensions, for a certain tuning of the coupling
constants, the Lovelock theory becomes equivalent to a Chern-Simons theory 
\cite{Chamseddine:1989nu}. Such theories have an enlarged local symmetry
group which allows (roughly speaking) to fit inside a \textquotedblleft
bigger curvature\textquotedblright\ for the AdS group both the standard
curvature and the torsion providing one with the needed mathematical
structures to formulate a supersymmetric theory as well. Because of such
local symmetry, the field equations are suitable to investigate non trivial
configurations such as black holes and worm holes (see, for instance, \cite%
{GOT07}). This combination of coefficients is unique in that the field
equations do not place strong constraints on the torsion. A black hole with
torsion was found in Ref. \cite{Ar06}. In that case, due to the enhanced
gauge symmetry of the Chern-Simons theory, the solution was related to a
torsion-free solution by a gauge transformation. Other solutions with
torsion in Chern-Simons gravity are given in Refs. \cite{Banados}.

In this paper we will exclude the Chern-Simons combination. The goal of this
paper will be to present some exact solutions with non-vanishing torsion in
five dimensional Lovelock gravity in the non-Chern-Simons case. It seems
that until now, no explicit solutions have been considered in the literature
(In Ref. \cite{Wheeler} there is a nice general discussion of spherically
symmetric torsion, where it was shown that static black holes in higher
dimensional Lovelock gravity generically have zero torsion). In order to
overcome the previously explained difficulties we use an ansatz of a
space-time being the cross product of the form $N_{2}\times M_{3}$ where
both submanifolds are of constant curvature and $M_{3}$ is spacelike. The
usefulness of this ansatz is due to the fact that one can search for a
torsion with components only in $M_{3}$. Because this submanifold is three
dimensional, a totally anti-symmetric torsion tensor $T_{ijk}\propto
\epsilon _{ijk}$ will respect the symmetry. Note also that such a torsion is
proportional to the Hodge dual of the curvature two form on $M_{3}$. In this
sense, the ansatz is inspired by an analogy with BPS states in gauge theory,
as will be explained later. The equations of motion impose a relation
between coupling constants which is \emph{not} the Chern-Simons combination.

One of the possible solutions is $AdS_{2}\times S_{3}$ which is analogous to
the Bertotti-Robinson solution \cite{Bertotti:1959pf} where the role of the
electromagnetic field is taken by a covariantly constant torsion. The
analogy to a BPS state is clear here as the Bertotti-Robinson solution is
indeed BPS. Because of this analogy one may wonder if also the solution
presented here is a BPS state. However there is no obvious way to write a
Killing spinor equation because, to the authors knowledge, the
supersymmetric extension of the theory is not known (although some features
of BPS states are indeed present as will be explained later). Anyway this
BPS analogy proved to be extremely useful to find nontrivial solutions with
torsion, which up to now has been an extremely difficult task.

It will be shown that this solution has no zero torsion limit. On the other
hand it is easy to see that the torsion free $AdS_2 \times S^3$ is a
solution for Einstein-Gauss-Bonnet gravity. The new solution presented here
seems to be therefore a topological excitation. It will be shown that also
other solutions with the cross product structure of constant curvature
manifolds exist.

The structure of the paper will be as follows: In section 2 the
Einstein-Gauss-Bonnet theory with torsion is reviewed. In section 3, the new
solutions are described. In section 4, the analogy with BPS states is
developed and some interesting features of the solutions are investigated.
Section 5 is a summary of the main conclusions.

\section{Einstein-Gauss-Bonnet theory with torsion}

\label{Results_Section}

\subsection{Gravity with Torsion}

Since the Kaluza-Klein idea and with the advent of string theories the
possibilities to have extra dimensions comes into play.

As mentioned previously, the Lovelock Lagrangian is the natural extension of
GR to higher dimensions. Let us briefly review Lovelock gravity in first
order formalism, the relation with the second order formalism will be
described in the next subsection in the five dimensional case. The action
has the form (for a detailed review, see Ref. \cite{TZ-CQG}): 
\begin{align*}
I_{D}& =\kappa \int \sum_{p=0}^{\left[ D/2\right] }\alpha _{p}L^{(D,p)}, \\
L^{(D,p)}& \equiv \varepsilon _{a_{1}\dots a_{D}}R^{a_{1}a_{2}}\wedge \cdots
\wedge R^{a_{2p-1}a_{2p}}\wedge e^{a_{2p+1}}\wedge \cdots \wedge e^{a_{D}}.
\end{align*}%
Here $e^{a}=e_{\mu }^{a}dx^{\mu }$ is the vielbein, $\omega _{\
b}^{a}=\omega _{b\mu }^{a}dx^{\mu }$ the spin connection, $\eta _{ab}$ the
Minkowski metric in the vielbein indices, $g_{\mu \nu }=\eta _{ab}e_{\mu
}^{a}e_{\nu }^{b}$ is the spacetime metric. The curvature and torsion are: 
\begin{align*}
R_{\ b}^{a}& \equiv \left( d\omega +\omega \wedge \omega \right) _{\ b}^{a},
\\
T^{a}& \equiv De^{a}=de^{a}+\omega _{\ b}^{a}\wedge e^{b}\ .
\end{align*}%
It is manifest in this way that the vielbein indices $a,b,..$ behave as
internal gauge indices. The Bianchi identities read%
\begin{align*}
DR_{\ b}^{a}& =dR_{\ b}^{a}+\omega _{\ c}^{a}\wedge R_{\ b}^{c}+\omega _{\
b}^{c}\wedge R_{\ c}^{a}=0, \\
DT^{a}& =R_{\ b}^{a}\wedge e^{b}.
\end{align*}

In $D=4$ the above tells that one is free to add a further term (the so
called Gauss-Bonnet term) to $S_{EH}$ which, being a topological invariant,
does not change the Euler-Lagrange equations. In higher dimensions the
situation changes: extremizing the Lovelock Lagrangian of order $p$ one
obtains the following types of equations: 
\begin{align}
\sum_{p=0}^{\left[ D/2\right] }(D-2p)\alpha _{p}\, \Xi _{a}^{(p)} &= 0\, , \\
\sum_{p=0}^{\left[ D/2\right] }p(D-2p)\alpha _{p}\, \Xi _{ab}^{(p)} & = 0\, ,
\end{align}
where 
\begin{align}
\Xi _{a}^{(p)}& \equiv\varepsilon
_{ab_{2}..b_{D}}R^{b_{2}b_{3}}..R^{b_{2p}b_{2p+1}}e^{b_{2p+2}} \dots
e^{b_{D}}\, ,  \label{loveq1} \\
\Xi _{ab}^{(p)}& \equiv\varepsilon
_{aba_{3}..a_{D}}R^{a_{3}a_{4}}..R^{a_{2p-1}a_{2p}}
T^{a_{2p+1}}e^{a_{2p+2}}\dots e^{a_{D}}\, .  \label{loveq2}
\end{align}

In four dimension $S_{EH}$ is basically the only first order action\footnote{%
It should be noted that there are more general actions with torsion that can
be constructed if one does not insist on a first order theory. In four
dimension the torsion plays an important geometrical role since the
important topological invariant (constructed by Nieh and Yan) can be
constructed $N=T^{a}T_{a}-e^{a}e^{b}R_{ab}$. Such an invariant appears in
the anomalous term of the divergence of the chiral anomaly \cite{CZ97}.}. In
this case the torsion plays no role, at least classically. In higher
dimensions the torsion emerges as a natural geometrical object.

\subsection{Five dimensional case}

In this paper, we consider the five dimensional Einstein-Gauss-Bonnet action
which in the familiar formalism reads:%
\begin{equation}
I=\kappa \int d^{5}x\sqrt{g}\left( R-2\Lambda +\alpha \left( R^{2}-4R_{\mu
\nu }R^{\mu \nu }+R_{\alpha \beta \gamma \delta }R^{\alpha \beta \gamma
\delta }\right) \right) \ ,  \label{Itensor}
\end{equation}%
where $\kappa $ is related to the Newton constant, $\Lambda $ to the
cosmological term, and $\alpha $ is the Gauss-Bonnet coupling. For later
convenience, it is useful to express the action (\ref{Itensor}) in terms of
differential forms as \footnote{%
The relationship between the constants appearing in Eqs (\ref{Itensor}) and (%
\ref{action}) is given by $\alpha =\frac{c_{2}}{2c_{1}}$, $\Lambda = -6\frac{%
c_{0}}{c_{1}}$, $\kappa =2c_{1}$.} 
\begin{equation}
I=\int \left( \frac{c_{0}}{5}e^{a}e^{b}e^{c}e^{d}e^{e}+\frac{c_{1}}{3}%
R^{ab}e^{c}e^{d}e^{e}+c_{2}R^{ab}R^{cd}e^{e}\right) \epsilon _{abcde}
\label{action}
\end{equation}%
where, as explained in the previous section, $e^{a}=e_{\mu }^{a}dx^{\mu }$
is the vielbein, and $R^{ab}=d\omega ^{ab}+\omega _{\text{ \ }f}^{a}\omega
^{fb}$ is the curvature $2$-form for the spin connection $\omega
^{ab}=\omega _{\ \mu }^{ab}dx^{\mu }$. The equations of motion obtained by
varying the action with respect to the spin connection $\omega ^{ab}$ read 
\begin{equation}
\mathcal{E}_{ab}\equiv T^{c}\left( c_{1}e^{d}e^{e}+2c_{2}R^{de}\right)
\epsilon _{abcde}=0  \label{equationtorsion}
\end{equation}%
The equations of motion obtained varying the action with respect to the
vielbein $e^{a}$ read 
\begin{equation}
\mathcal{E}_{e}\equiv \left(
c_{o}e^{a}e^{b}e^{c}e^{d}+c_{1}R^{ab}e^{c}e^{d}+c_{2}R^{ab}R^{cd}\right)
\epsilon _{abcde}=0.  \label{equationcurvature}
\end{equation}%
When the coefficients of the theory satisfy a special fine-tuning the above
action turns out to be of a Chern-Simons theory\cite{Chamseddine:1989nu}. In
this section the case of Chern-Simons will be explicitly excluded by
imposing the inequality: 
\begin{equation}
c_{1}^{2}\neq 4c_{0}c_{2}\ .
\end{equation}%
or, equivalently 
\begin{equation*}
\frac{4\alpha \Lambda }{3}\neq -1.
\end{equation*}

\section{Exact solutions with torsion}

The main idea is that an ansatz, inspired by BPS states in field theory,
could allow one to find non trivial solutions with non vanishing torsion. As
BPS states in field theory have non trivial topological charges, in the
present case such \textit{vacuum\ solutions} of Einstein-Gauss-Bonnet
gravity could manifest some properties of solutions in the presence of
matter fields carrying some charges. For a suitable choice of the
coefficients which \textit{is not} Chern-Simons, a set of solutions will be
constructed.

One of these solutions is the vacuum analogue of the Bertotti-Robinson
metric in which the torsion plays the role of the electromagnetic field.
This will be derived in the next subsection before proceeding to the more
general solutions.

\subsection{$AdS_{2}\times S_{3}$ solution}

\label{AdS_Section}

We search for a $AdS_{2}\times S_{3}$ solution with torsion. The idea is to
find the analogous of a Bertotti-Robinson metric in which the torsion plays
the role of the electromagnetic field. Therefore the following ansatz for
the metric is natural 
\begin{gather}
ds^2 =\frac{l^2}{x^2} \left( - dt^2 + dx^2\right) + \frac{r_0^2}{4}\left(
d\phi^2 + d\theta^2 + d\psi^2 + 2 \cos\theta d\phi d\psi\right)\, .
\end{gather}
As vielbein we choose: 
\begin{equation}
e^{0}=\frac{l}{x}dt\;\;\;;\;\;\;e^{1}=\frac{l}{x}dx\;\;\;;\;\;\;e^{i}= r_0%
\tilde{e}^{i}  \label{1ansatz1}
\end{equation}%
where $r_0$, the radius of the 3-sphere, is a constant and $\tilde{e}^{i}$
is the intrinsic vielbein on the unit sphere. For definiteness, the Poincar%
\'{e} coordinates have been used for the two dimensional Anti de Sitter
space and Euler angles for the sphere.

We make the following ansatz for the torsion, consistent with the spherical
symmetry: 
\begin{equation}
T^{1}=0\;\;\;;\;\;\;T^{0}=0\;\;\;;\;\;\;T_{i}=\frac{H}{r_0}%
e^{j}e^{k}\epsilon _{ijk}\, ,  \label{1torsion1}
\end{equation}
where $H$ is a constant.

Note that on the unit three-sphere there exists a choice of intrinsic
vielbein such that $\tilde{\omega}^{ij}=-\epsilon ^{ijk}\tilde{e}_{k}$,
where $\tilde{\omega}^{ij}$ is the intrinsic Levi-Civita spin connection.
With this choice, the 5-dimensional spin connection reads 
\begin{equation}
\omega^{01}=-\frac{1}{x}dt\, , \qquad \omega^{ij} =(H+1)\tilde{\omega}^{ij}
= - (H+1)\epsilon ^{ijk}\tilde{e}_{k} \, .  \label{2ansatz2}
\end{equation}

Thanks to the geometric structure of the sphere, the torsion can be written
in a way that is homogeneous and isotropic thanks to the invariance of the
tensor $\epsilon ^{ijk}$ on the sphere. The naturalness of this ansatz will
be discussed more in section 4. The curvature turns out to be 
\begin{equation}
R^{01}=-\frac{1}{l^{2}}e^{0}e^{1};\;\;\;R^{ij}=\frac{1-H^{2}}{r_0^{2}}%
\,e^{i}e^{j}  \label{1curvature1}
\end{equation}%
One recovers the torsionless case setting $H=0$. Inserting Eqs. (\ref%
{1ansatz1}), (\ref{1torsion1}) and (\ref{1curvature1}) in the equations of
motion one obtains from the $(ij)$ component of (\ref{equationtorsion}): 
\begin{equation}
c_{1}-\frac{2c_{2}}{l^{2}}=0\qquad \mathrm{or}\qquad H=0\,.
\label{AdSxS_Solution_1}
\end{equation}%
The other components of equation (\ref{equationtorsion}) are automatically
satisfied. From the $(0)$ and $(1)$ components of eq. (\ref%
{equationcurvature}) we get 
\begin{equation}
4c_{0}+2c_{1}\frac{(1-H^{2})}{r_0^{2}}=0  \label{AdSxS_Solution_2}
\end{equation}%
The $(i)$ component of eq. (\ref{equationcurvature}) gives 
\begin{equation}
12c_{0}+2c_{1}\left( -\frac{1}{l^{2}}+\frac{1-H^{2}}{r_0^{2}}\right) -\frac{%
4c_{2}}{l^{2}}\frac{(1-H^{2})}{r_0^{2}}=0  \label{AdSxS_Solution_3}
\end{equation}

It is worth stressing here that the form of the torsion of Eq. (\ref%
{1torsion1}) greatly simplify the equations of motions (which in the
Einstein-Gauss-Bonnet are quite complicated). In particular, some key
identities have been used in deriving the above equations. The first one is
that expressions like 
\begin{equation}
\epsilon _{ijk}T^{i}e^{j}e^{k}=0  \label{1I1}
\end{equation}%
identically vanish due to the fact that the torsion contains always two
angular vielbeins so that such exterior products are zero because there are
only three independent angular vielbeins. The second identity is that
expressions like%
\begin{equation}
\epsilon _{ijk}T^{i}e^{j}e^{0}\approx \epsilon _{ijk}\left( \epsilon
^{imn}e_{m}e_{n}\right) e^{j}e^{0}=\left( \delta _{j}^{m}\delta
_{k}^{n}-\delta _{k}^{m}\delta _{j}^{n}\right) e_{m}e_{n}e^{j}e^{0}=0
\label{2I2}
\end{equation}%
also vanish since there is always a wedge product of an angular vielbein
with itself.

Substituting equations (\ref{AdSxS_Solution_1}) and (\ref{AdSxS_Solution_2})
in (\ref{AdSxS_Solution_3}) one finds that there exist solutions with
torsion only if the coupling constant satisfy the following relation 
\begin{equation}
c_{1}^{2}=12c_{0}c_{2}  \label{nocs}
\end{equation}%
Note that this is \emph{not} the Chern-Simons combination of the coupling
constants. The $AdS_2$ length scale $l$ is completely determined by the
coupling coefficients: 
\begin{equation}  \label{Ads_length}
\frac{1}{l^2} = \frac{c_1}{2c_2}\, ,
\end{equation}
The sphere radius $r_0$ and the torsion parameter $H$ are related by 
\begin{equation*}
1- H^2 = - \frac{2c_{0}}{c_{1}}r_0^{2}\quad \Rightarrow \quad H^2 = 1 + 
\frac{r_0^2}{3l^2}\, .
\end{equation*}
In summary, the space-time $AdS_2\times S_3$ with vielbein given by (\ref%
{1ansatz1}) and with torsion and curvature 
\begin{gather}
T_{i}=\pm \sqrt{\frac{1}{r_0^2} + \frac{1}{3l^2}}\ \epsilon _{ijk}\,
e^{j}e^{k}\, ,  \label{Answer_Torsion} \\
R^{01}=-\frac{1}{l^2} e^{0}e^{1}\,, \qquad R^{ij}=-\frac{1}{3l^{2}}%
\,e^{i}e^{j}\, ,
\end{gather}
with $l$ given by (\ref{Ads_length}), is a solution of the
Einstein-Gauss-Bonnet theory, provided that the relation (\ref{nocs}) among
the coupling constants holds, with $c_2/c_1$ and $c_0/c_1$ positive.

The torsion is bounded from below for finite sized sphere and AdS length
scale. This means that there is no continuous zero torsion limit.

Moreover, the torsion is fully antisymmetric. This allows an intriguing
analogy with gravity in the presence of a constant electromagnetic field. It
is natural to define a three form $T \equiv T_{ijk}e^i\wedge e^j \wedge e^k
= 3! \frac{H}{r_0} \, e^2\wedge e^3 \wedge e^4$. Also we define the Hodge
dual, the two form $*T = - 3! \frac{H}{r_0} e^0\wedge e^1$. Due to the
Bianchi identities, these are covariantly constant. 
\begin{equation*}
DT = 0,\qquad D *T =0.
\end{equation*}
The analogy with electromagnetic field is made by defining $F \equiv *T$.
Thus $F$ is seen to obey the source-free Maxwell equations, making manifest
the close resemblance with the electromagnetic field of the
Bertotti-Robinson solution.

\subsection{Product of two constant curvature manifolds with torsion}

In the previous section a product manifold was considered. It was seen that
the torsion could be introduced on the three-sphere in a way that was
consistent with spherical symmetry. Furthermore such a torsion satisfied the
equations of motion, provided that there was a rather curious relation (\ref%
{nocs}) between the coupling constants in the action. A key feature of that
solution was that the torsion tensor $T_{ijk} \propto \epsilon_{ijk}$ is
manifestly of a form which is homogeneous and isotropic in the
three-dimensional subspace (in that case a sphere). It is natural to
generalise this solution to more general product manifolds involving a three
dimensional manifold with constant curvature, which could be positive,
negative or zero. In this section we shall study solutions whose metric is
the direct product of two manifolds of constant Riemannian curvature, with
torsion living in a three-dimensional manifold.

Let $N_2$ be a two-dimensional manifold with Minkowskian signature and
constant curvature. The metric is given by $ds^2_N = - e^0 \otimes e^0 + e^1
\otimes e^1$, where $e^0$ and $e^1$ are the vielbeins with Levi-Civita
connection $\hat{\omega}^{01}$ and curvature satisfying 
\begin{gather}  \label{curvature_01}
\hat{R}^{01} = \Lambda_N\, e^0 \wedge e^1\, .
\end{gather}
Let $M_3$ be a three-dimensional manifold of constant curvature with
Euclidean signature. The metric is $ds^2_M = \delta_{ij}\, e^i \otimes e^j$.
The three-manifold has a Levi-Civita connection\footnote{%
Since torsion shall be later introduced on $M_3$, the Levi-Civita connection
shall be denoted by $\hat{\omega}^{ij}$. The symbol $\omega^{ij}$ is
reserved for the full connection including the contorsion.} $\hat{\omega}%
^{ij}$ and corresponding curvature $\hat{R}^{ij}$ satisfying 
\begin{gather}  \label{curvature_ij}
\hat{R}^{ij} = \Lambda_M\, e^i \wedge e^j\, .
\end{gather}

The five-dimensional spacetime will be a product space $N_2 \times M_3$, the
metric being 
\begin{gather}
ds^2 = ds^2_N + ds^2_M\,.
\end{gather}
The torsion is introduced onto the $M_3$ in such a way as to respect the
symmetry. This is guaranteed by the invariance property of the alternating
tensor $\epsilon_{ijk}$. Let us make the ansatz: 
\begin{gather}  \label{symmetric_Torsion}
T_i = \tau \, \epsilon_{ijk} e^j \wedge e^k\, .
\end{gather}
We shall further assume that $\tau$ is constant, which implies that $T^i$ is
covariantly constant with respect to the Levi-Civita connection. In the
appendix a more general ansatz is analysed where the symmetry of $N_2$ is
relaxed and it is shown that the only solution is the one discussed here.

The spin connection on $M_3$ now takes the form $\hat{\omega}^{ij} + K^{ij}$
where $K^{ij}$ is the contorsion 1-form. According to (\ref%
{symmetric_Torsion}) the contorsion is: 
\begin{gather}  \label{Symmetric_contorsion}
K_{ij} = -\tau\, \epsilon_{ijk} e^k\, .
\end{gather}

Let $R^{ab}$ denote the five-dimensional curvature tensor. It is a sum of
the torsion-free curvature, with components (\ref{curvature_01}) and (\ref%
{curvature_ij}), and a part which comes from the torsion. This can
conveniently be obtained by expanding $R^{ab}=d(\hat{\omega}^{ab}+K^{ab})+(%
\hat{\omega}_{\ c}^{a}+K_{\,c}^{a})\wedge (\hat{\omega}^{cb}+K^{cb})$ to
give the well-known formula: 
\begin{equation}  \label{well_known}
R^{ab}=\hat{R}^{ab}+\hat{D}K^{ab}+K_{\ c}^{a}\wedge K^{cb}\,.
\end{equation}
\ \ From (\ref{Symmetric_contorsion}) it can be seen that the contorsion is
covariantly constant with respect to the Levi-Civita connection, $\hat{D}%
K^{ij}=\kappa \left( \epsilon _{ljk}\hat{\omega}_{\ i}^{l}+\epsilon _{ilk}%
\hat{\omega}_{\ j}^{l}+\epsilon _{ijl}\hat{\omega}_{\ k}^{l}\right) \wedge
e^{k}=0$. The non-vanishing components of the curvature are 
\begin{equation}  \label{Symmetric_Curvature}
R^{01}=\Lambda _{N}e^{0}\wedge e^{1}\,,\qquad R^{ij}=\left( \Lambda
_{M}-\tau ^{2}\right) e^{i}\wedge e^{j}\,.
\end{equation}%
The effect of the homogeneous and isotropic torsion is to rescale the
three-dimensional part of the curvature.

Now it remains to check that the field equations are satisfied by torsion (%
\ref{symmetric_Torsion}) and curvature (\ref{Symmetric_Curvature}). First
let us check equation (\ref{equationtorsion}) coming from the variation
w.r.t. the spin connection. Since $\epsilon_{ijk} T^j \wedge e^k = 0$ the
only non-trivial component is: 
\begin{gather}  \label{dev1}
0 = \mathcal{E}_{ij} = 2\epsilon_{ijk} T^k \wedge \left(c_1 e^0 \wedge e^1 +
2c_2 R^{01}\right)\, .
\end{gather}
Now we check equation (\ref{equationcurvature}) coming from the variation
w.r.t. the vielbein. 
\begin{align}
0 = \mathcal{E}_0 & = \epsilon_{ijk} \left( 4 c_0 e^1 \wedge e^i \wedge e^j
\wedge e^k + 2 c_1 e^1 \wedge e^i \wedge R^{jk}\right)\, ,  \label{dev2} \\
0 = \mathcal{E}_1 & = \epsilon_{ijk} \left( 4 c_0 e^0 \wedge e^i \wedge e^j
\wedge e^k + 2 c_1 e^0 \wedge e^i \wedge R^{jk}\right)\, ,  \label{dev3} \\
0 = \mathcal{E}_i & = \epsilon_{ijk} \Big( 12 c_0 e^0 \wedge e^1 \wedge e^j
\wedge e^k  \notag \\
&\qquad\quad + 2 c_1 ( R^{01} \wedge e^j \wedge e^k + e^0\wedge e^1 \wedge
R^{jk}) + 4 c_2 R^{01}\wedge R^{ij} \Big)\, .  \label{dev4}
\end{align}
Equation (\ref{dev1}) implies: 
\begin{gather}
\Lambda_N = -\frac{c_1}{2c_2}\, .
\end{gather}
Equations (\ref{dev2}) and (\ref{dev3}) imply 
\begin{gather}
\Lambda_M - \tau^2 + \frac{2c_0}{c_1} = 0\, .
\end{gather}
Substituting these two equations in (\ref{dev4}) gives the relation between
the coupling constants $c_1^2 = 12 c_0c_2$.

\subsection{Summary of the solutions}

We have found solutions for the special class of Lovelock theories satisfying%
\footnote{%
In terms of the notation of equation (\ref{Itensor}), the coefficients
satisfy $4\alpha \Lambda =-1$.} $c_1^2 = 12 c_0c_2$. Since the possibility
of a vanishing Einstein-Hilbert term is excluded, we may normalise $c_1 = 1$
without loss of generality. The action is thus: 
\begin{gather}  \label{special_action}
I=\int \left( \frac{1}{60c_2}e^{a}e^{b}e^{c}e^{d}e^{e}+\frac{1}{3}%
R^{ab}e^{c}e^{d}e^{e}+c_{2}R^{ab}R^{cd}e^{e}\right) \epsilon _{abcde}\, .
\end{gather}

The metric is the product of $N_2 \times M_3$ with Riemannian curvature: 
\begin{gather}
\hat{R}^{01} = - \frac{1}{2c_2} e^0 \wedge e^1\, , \quad \hat{R}^{ij} =
\Lambda_M e^i \wedge e^j\, .
\end{gather}
The curvature and torsion are: 
\begin{gather}
R^{01} = - \frac{1}{2c_2} e^0 \wedge e^1\, , \quad R^{ij} = - \frac{1}{6c_2}
e^i \wedge e^j\, , \\
T_i = \pm \sqrt{ \Lambda_M + \frac{1}{6c_2} } \,\, \epsilon_{ijk}\, e^j
\wedge e^k\, .
\end{gather}
We see that the full non-Riemannian curvature is completely determined by
the coupling constant $c_2$. There is just one constant of integration $%
\Lambda_M$, which characterizes both the Riemannian curvature of $M_3$ and
the torsion. \newline

\begin{center}
\begin{tabular}{|c|c|c|c|}
\hline
$\mathbb{R}_2 \times M_3$ &  & No solution &  \\ 
AdS$_2 \times S_3$ & $c_2 > 0$ & $\Lambda_M$ arbitrary & No zero torsion
limit \\ 
AdS$_2 \times H_3$ & $c_2 > 0$ & $- 1/6c_2 \leq \Lambda_M <0$ & Zero torsion
limit \\ 
AdS$_2 \times \mathbb{R}_3$ & $c_2 > 0$ & $\Lambda_M = 0$ & No zero torsion
limit \\ 
dS$_2 \times S_3$ & $c_2 < 0$ & $1/6|c_2| \leq \Lambda_M $ & Zero torsion
limit \\ 
dS$_2 \times H_3$ & ($c_2 < 0$) & No solution &  \\ 
dS$_2 \times \mathbb{R}_3$ & ($c_2 < 0$) & No solution &  \\ \hline
\end{tabular}
\end{center}

Note that the generalisation to the case that $N_2$ is Euclidean and with $%
M_3$ Lorentzian is straightforward.

The field equations for Einstein-Gauss-Bonnet can also be written as
follows: 
\begin{align}
& T^{c}\left( R^{de}+\frac{(\Lambda _{+}+\Lambda _{-})}{2}e^{d}e^{e}\right)
\epsilon _{abcde}=0  \label{eto1} \\
& \left( R^{ab}+\Lambda _{+}e^{a}e^{b}\right) \left( R^{cd}+\Lambda
_{-}e^{c}e^{d}\right) \epsilon _{abcde}=0  \label{eto2}
\end{align}%
so that, assuming that the torsion vanishes there are two possible (A)dS
vacua with different cosmological constant. Here we have found a third kind
of solution with a high degree of symmetry: the product of maximally
symmetric spaces with nonzero torsion. In the solutions we have found above,
it is the average of the two cosmological constants which is important, 
\begin{equation*}
\pm \frac{1}{l^{2}}=\frac{(\Lambda _{+}+\Lambda _{-})}{2},
\end{equation*}
because it determines the cosmological constant of the $N_2$.

\section{Field theoretical features in first-order gravitational theories}

In Ref. \cite{Ca07} some analogies were investigated between BPS states in
field theory on the one hand and Gravitational theories with torsion on the
other. Let us briefly revisit this subject in the light of the solutions
found in section \ref{Results_Section}. For detailed reviews on BPS states
in field theory see Refs. \cite{OW99,To05}. We have focused on the Lovelock
theories because they have a first order formalism. It is not a scope of the
present paper to analyze all the possible higher curvature corrections
(which are expected on various theoretical grounds ranging from string
theory to Kaluza-Klein reductions) to standard Einstein-Hilbert action,
because such corrections are higher order in derivatives. A detailed review
on how generic higher curvature corrections may arise and on their
interesting physical effects can be found in \cite{Sch06} and references
therein. On the other hand, the formal analogy pointed out in \cite{Ca07} is
only based on the geometrical roles of the Higgs field and the gauge
connection on one hand and of the vielbein and the spin connection on the
other hand while the detailed form of the field equations is not so
important. Therefore, the present approach to analyze the dynamical role of
torsion could also work in some more general cases not belonging to the
Lovelock class.

In the Yang-Mills-Higgs theory, the BPS equations involve typically linear
relations (such as higher dimensional self-duality conditions) among $%
D^{a}\phi $ [the covariant derivative of the Higgs field] and $F^{ab}$ [the
Yang-Mills field strength] in which $\phi $ can enter quadratically (as, for
instance, in the vortex case\cite{OW99,To05}). Inspired by this, a natural
ansatz for gravity with torsion is the linear relation 
\begin{equation}
T^{c}=f_{ab}^{c}\left( \alpha R^{ab}+\beta e^{a}e^{b}\right)
\label{gravBPS1}
\end{equation}%
where $f_{ab}^{c}$ is an appropriately chosen three index tensor and $\alpha 
$ and $\beta $ are two constants.

Now there does not exist a genuine invariant tensor with three indices (that
is, $f_{ab}^{c}$ in Eq. (\ref{gravBPS1})) in the five-dimensional Lorentz
group. A solution involving such a tensor would necessarily break some of
the Lorentz symmetry. One of the features of topological defects is
precisely that they partially break Lorentz invariance (the surviving
generators are the ones leaving invariant the defects). For instance, when
in quantum field theory one expands around the trivial vacuum (all the
fields equal to zero) the Lorentz generators annihilate the vacuum (see, for
instance, \cite{We96}). When expanding around non trivial saddle points
(that is, topological defects) this is not so since the position and the
structure of the topological defects make the action of the Lorentz
generators on the vacuum non trivial.

In the case of our solutions, there is a three-dimensional submanifold of
constant curvature. Thus it is natural to choose the tensor $%
f_{ijk}=\epsilon _{ijk}$ consistent with the unbroken Lorentz generators.

Since our solutions are of the form (\ref{gravBPS1}), there is some analogy
between our solutions an BPS states. However, because of the very different
structure of Lovelock action, it is not easy to make this analogy precise.
It is not easy to construct an energy functional from which to deduce a BPS
bound. In spite of the fact that the supergravity or a BPS bound are not
known, some interesting features of BPS states can be analyzed without the
tools of SUSY. In particular, in the vacuum analogue of the
Bertotti-Robinson it appears that the torsion is bounded from below for
finite sized sphere and AdS length scale. This means that there is no
continuous zero torsion limit.\ So that such a solution appears as a
topological excitation in which intrinsically the torsion \textit{cannot be
small}. Another interesting feature is related to the rigidity of the
solutions due to the presence of torsion. In the non Chern-Simons case, the
equations of motions have to fulfil a strong compatibility condition: taking
the covariant derivative of Eq. (\ref{equationcurvature}) and comparing with
Eq. (\ref{equationtorsion}) it turns out that \cite{Troncoso:1999pk} 
\begin{eqnarray*}
e^{b}R^{cd}T^{e}\epsilon _{abcde} &=&0, \\
e^{b}e^{c}e^{d}T^{e}\epsilon _{abcde} &=&0.
\end{eqnarray*}%
Such conditions are absent in the Chern-Simons case because of the tuning of
the coefficients. The rigidity of the above conditions makes clear why it is
so difficult to find exact solutions with intrinsic torsion (when the
torsion is absent the above conditions are trivial). It is then apparent
that the ``BPS-inspired" ansatz for the torsion proposed in \cite{Ca07} is
quite good because it naturally provides one with a method to solve the
above conditions due to the identities (\ref{1I1}) and (\ref{2I2}).

As further evidence that the solution may be a topologically non-trivial
vacuum, it is worth pointing out that the action (\ref{special_action}),
evaluated on the solutions is zero (compare with Euclidean instantons which
have finite action), as can be easily checked. In contrast, the action
evaluated on the two $AdS_5$ solutions, does not vanish.

\section{Conclusions and outlook}

It has been shown that in five dimensional Einstein-Gauss-Bonnet gravity
there are exact solutions with non vanishing intrinsic torsion. These were
found for a special choice of coupling constants given by the action (\ref%
{special_action}), which \emph{is not the Chern-Simons combination}. To the
best of the authors knowledge, no solutions with torsion in non Chern-Simons
Einstein-Gauss-Bonnet gravity have been found before. The analogies with
field theory and the peculiar features of such states have been discussed.
Among the solutions found there is an analogue of the Bertotti-Robinson
metric in which the torsion is the dual of a covariantly constant
electromagnetic field. However, without knowing the supersymmetric extension
of this theory, there is no obvious way to write the Killing spinor
equation. This solution is, in a sense a topological excitation in which
there is no continuous zero torsion limit. Because of the rigidity of the
Einstein-Gauss-Bonnet equations of motions in the presence of nontrivial
torsion there are strong constraints on small excitations (besides the
re-scaling of the physical parameters appearing in the solutions). It is
reasonable to expect that, for this reason, the above exact solutions with
the ``BPS-inspired" torsion could be a topologically non-trivial vacuum
state.

In view of the analogy with the Bertotti-Robinson solution, one may also
expect the existence of a solution analogue of the extreme
AdS-Reissner-Nordstrom black hole which interpolates between the
Bertotti-Robinson and the AdS metric. This is an open question for future
research.

\bigskip

\section*{Acknowledgements}

The authors would like to thank Ricardo Troncoso for many stimulating
discussions and important bibliographic suggestions. We also thank E.
Gravanis, H. Maeda, J. Oliva and J. Zanelli for many suggestions and
continuous encouraging. This work has been partially supported by PRIN
SINTESI 2007 and by Proy. FONDECYT N%
${{}^\circ}$%
3070055, 3070057, 3060016 and by institutional grants to CECS of the
Millennium Science Initiative, Chile, and Fundaci\`{o}n Andes, and also
benefits from the generous support to CECS by Empresas CMPC.

\appendix

\section{On the generality of the ansatz}

Let us now argue that the solutions found in the previous section are quite
general for a product metric involving a maximally symmetric three-manifold $%
M_3$. Let us now assume that the metric is $N_2\times M_3$, where now $N_2$
is an arbitrary two-manifold. That is to say, we shall not specify the form
of $\hat{R}^{01}$.

We shall keep the same ansatz for the torsion on $M_3$ except that, since $%
N_2$ need not be maximally symmetric, $\tau$ may depend on $(x,t)$, the
co-ordinates of $N_2$. Thus we look for solutions of the form: 
\begin{gather}
T^0 = F(x,t)\, e^0 \wedge e^1\, , \quad T^1 = G(x,t)\, e^0 \wedge e^1\,
,\quad T_i = \tau(x,t)\, \epsilon_{ijk} e^j \wedge e^k\, .
\end{gather}
The contorsion is: 
\begin{gather*}
K^{01} = F e^0 - G e^1\, , \qquad K_{ij} = -\tau\, \epsilon_{ijk} e^k\, ,
\end{gather*}
The components of curvature along $M_3$ are found using the formula (\ref%
{well_known}). 
\begin{gather}
R^{ij} = \left(\Lambda_M - \tau^2 \right) e^i e^j - d\tau \,\epsilon^{ijk}\,
e_k\, .
\end{gather}

Let us study the field equations (\ref{equationtorsion}) and (\ref%
{equationcurvature}). The component $\mathcal{E}_{01} =0$ tells us
immediately that $d\tau$ vanishes: 
\begin{gather}
\tau = \text{constant}\, .
\end{gather}
The components $\mathcal{E}_{0i} =0$ and $\mathcal{E}_{1i} =0$ imply: 
\begin{gather}  \label{contradiction1}
1-\tau^2 +\frac{c_1}{2c_2} =0\qquad \text{or} \qquad F =0,\ G = 0\ .
\end{gather}
The component $\mathcal{E}_{0} = 0$ gives: 
\begin{gather}  \label{contradiction2}
1-\tau^2 + \frac{2 c_0}{c_1} =0\, .
\end{gather}
Comparing equations (\ref{contradiction1}) and (\ref{contradiction2}) we
have either $4c_0c_2 = c_1^2$ or $F = G=0$. The first of these alternatives
is precisely the Chern-Simons combination. So we conclude that, for $4c_0c_2
\neq c_1^2$, we have $F = G = 0$ and $\tau = $ constant.

Finally, $\mathcal{E}_{ij} =0$ imposes that the curvature of $N_2$ is
constant. Thus the solution reduces to that of the previous section.

\end{document}